\documentclass[sigconf]{acmart}

\AtBeginDocument{%
  \providecommand\BibTeX{{%
    \normalfont B\kern-0.5em{\scshape i\kern-0.25em b}\kern-0.8em\TeX}}}

\copyrightyear{2024}
\acmYear{2024}
\setcopyright{acmlicensed}\acmConference[SIGIR '24]{Proceedings of the 47th International ACM SIGIR Conference on Research and Development in Information Retrieval}{July 14--18, 2024}{Washington, DC, USA}
\acmBooktitle{Proceedings of the 47th International ACM SIGIR Conference on Research and Development in Information Retrieval (SIGIR '24), July 14--18, 2024, Washington, DC, USA}
\acmDOI{10.1145/3626772.3657963}
\acmISBN{979-8-4007-0431-4/24/07}

\begin{document}

\title{USimAgent: Large Language Models for Simulating Search Users}


\author{Erhan Zhang}
\affiliation{%
  \institution{GSAI, Renmin University of China}
  \city{Beijing}
  \country{China}}
\email{erhanzhang@ruc.edu.cn}

\author{Xingzhu	Wang}
\affiliation{%
  \institution{GSAI, Renmin University of China}
  \city{Beijing}
  \country{China}}
\email{wangxingzhu2022@ruc.edu.cn}

\author{Peiyuan	Gong}
\affiliation{%
  \institution{GSAI, Renmin University of China}
  \city{Beijing}
  \country{China}}
\email{pygongnlp@gmail.com}

\author{Yankai Lin}
\affiliation{%
  \institution{GSAI, Renmin University of China}
  \city{Beijing}
  \country{China}}
\email{yankailin@ruc.edu.cn}

\author{Jiaxin Mao}
\authornote{Corresponding author.}
\affiliation{%
  \institution{GSAI, Renmin University of China}
  \city{Beijing}
  \country{China}}
\email{maojiaxin@gmail.com}



\begin{abstract}
Due to the advantages in the cost-efficiency and reproducibility, user simulation has become a promising solution to the user-centric evaluation of information retrieval systems. Nonetheless, accurately simulating user search behaviors has long been a challenge, because users' actions in search are highly complex and driven by intricate cognitive processes such as learning, reasoning, and planning. Recently, Large Language Models (LLMs) have demonstrated remarked potential in simulating human-level intelligence and have been used in building autonomous agents for various tasks. However, the potential of using LLMs in simulating search behaviors has not yet been fully explored. In this paper, we introduce a LLM-based \textbf{u}ser search behavior \textbf{sim}ulator, \textbf{USimAgent}. The proposed simulator can simulate users' querying, clicking, and stopping behaviors during search, and thus, is capable of generating complete search sessions for specific search tasks. Empirical investigation on a real user behavior dataset shows that the proposed simulator outperforms existing methods in query generation and is comparable to traditional methods in predicting user clicks and stopping behaviors. These results not only validate the effectiveness of using LLMs for user simulation but also shed light on the development of a more robust and generic user simulators. The code and data are accessible at \url{https://github.com/Meow-E/USimAgent}.
\end{abstract}





\maketitle

\vspace{-9pt}
\section{Introduction}

Recently, user simulation has become a promising solution to the user-centric evaluation of information retrieval systems~\cite{balog2023user}. By generating a vast amount of data through simulators without the need for actual user involvement, user simulation offers not only cost-effectiveness but also significant advantages in terms of flexibility and efficiency in data collection. Additionally, simulators are not prone to issues like fatigue from prolonged search activities, ensuring the consistency and repeatability of experimental results.

Given the complexity of user search behavior, simulating the whole search session has long been a challenging and enduring task. Traditional simulation methods typically decompose a user's interactive search behavior into a series of independent steps, including submitting queries, browsing Search Engine Results Pages (SERPs), clicking results, reading and evaluating documents, and deciding when to stop~\cite{balog2023user}. Therefore, they require a dedicated simulation strategy to be designed for each step, such as generating search queries by extracting terms from language models associated with specific documents or topics\cite{DBLP:conf/sigir/AzzopardiRB07, DBLP:conf/sigir/Azzopardi09, DBLP:conf/jcdl/JordanWG06, DBLP:conf/sigir/BaskayaKJ12, DBLP:conf/ictir/CarteretteBZ15}, estimating the probability of a user clicking on search results based on historical data using click models\cite{DBLP:conf/wsdm/CraswellZTR08, DBLP:conf/wsdm/GuoLW09, DBLP:conf/www/ChapelleZ09, DBLP:conf/sigir/DupretP08, DBLP:conf/www/BorisovMRS16}, and deciding when a user stops searching based on a set of predefined simplistic assumptions through heuristic rules\cite{DBLP:journals/jasis/Cooper73a, DBLP:journals/ipm/KraftL79, 1995Judgment, DBLP:conf/cikm/MaxwellAJK15, DBLP:journals/sigir/Maxwell19}. However, these approaches fail to fully consider the dynamic and interdependent nature of user behavior. In particular, they often fail to model how some cognitive factors, such as the information needs and background knowledge, and cognitive processes, including the learning and reasoning, would affect and drive user's action. These limitations make it difficult for the existing user simulation methods to accurately and comprehensively reproduce the complex interactions between users and search engines.

With an astonishing number of parameters and vast amount of text knowledge, Large Language Models (LLMs) have been substantially enhanced and achieved human-level intelligence in many tasks\cite{DBLP:conf/nips/Yao0YN22, DBLP:journals/corr/abs-2302-04761, DBLP:journals/corr/abs-2206-08853, DBLP:journals/corr/abs-2305-17390}. LLMs demonstrate considerable potential in simulating user behavior, primarily reflected in the following aspects: (1) LLMs possess the ability to understand and process natural language\cite{DBLP:conf/nips/Ouyang0JAWMZASR22}, meaning that they can accurately comprehend task instructions and user query intentions; (2) LLMs exhibit strong Zero-shot/Few-shot learning capabilities\cite{DBLP:conf/nips/BrownMRSKDNSSAA20, DBLP:journals/corr/abs-2204-02311, DBLP:conf/nips/KojimaGRMI22}, thereby offering greater flexibility to adapt to and address the diverse scenarios and requirements encountered in simulating user behavior; (3) Multi-task fine-tuning enables LLM-based agents to achieve outstanding performance across multiple tasks and domains, which allows a single model to simulate a wide range of user interaction behaviors; (4) LLMs have demonstrated the capacity for logical reasoning, decision-making, and planning multi-step tasks within a given context\cite{DBLP:conf/iclr/YaoZYDSN023, DBLP:journals/corr/abs-2310-04406}. This trait can enable LLMs to maintain logical coherence in complex interactions, making the simulated user behavior more closely resemble the coherent thinking and search patterns of users in the real world.

Having said that, how to use LLMs in simulating search behavior has not yet been fully investigated. Therefore, in this paper, we propose \textbf{USimAgent}, a new framework that leverages LLMs for search \textbf{u}ser \textbf{sim}ulation. Specifically, when given a search task, USimAgent simulates users' search behavior by continuously alternating between generating queries and clicks, until a stopping action is generated, which ultimately reproduce a complete interaction sequence for the task. Leveraging LLMs' ability in language understanding, USimAgent is capable of integrating information from the session context and external environment to enhance the realism of the simulation. Powered by the zero-shot/few-shot capabilities, it can flexibly adapt to different scenarios without the need for additional training for each search task. Inspired by ReAct\cite{DBLP:conf/iclr/YaoZYDSN023}, USimAgent expands the action space to include combinations of possible reasoning and action steps, which further allows it to engage in in-depth reasoning based on the current context before executing actions, thereby producing coherent behavioral outputs. 

To evaluate the proposed framework and compare it against existing user simulation approaches, we conduct experiments on a public user behavior dataset\cite{DBLP:conf/kdd/LiuMLZM19}. This dataset includes everyday information retrieval and decision-making tasks, aiming to simulate the regular search behaviors of humans on the internet. We analyzed the interaction quality generated by USimAgent by comparing it with real user behavior data. The experimental results indicate that in terms of query generation, our LLM-based agent outperforms existing methods. In simulating clicking and stopping behavior, our method is comparable to traditional models. These findings not only demonstrate the feasibility of using LLMs for user simulation but also highlight the necessity of further developing more robust intelligent agents.

\vspace{-8pt}
\section{Related Work}

\noindent \textbf{User simulation.} Numerous studies have focused on simulating user interactions with search engines. Simulating queries is one of the most challenging aspects of user simulation. Most approaches construct queries by sampling terms from language models associated with specific documents or topics\cite{DBLP:conf/sigir/AzzopardiRB07, DBLP:conf/sigir/Azzopardi09, DBLP:conf/jcdl/JordanWG06, DBLP:conf/sigir/BaskayaKJ12, DBLP:conf/ictir/CarteretteBZ15}. Click models are widely used to simulate user clicking behavior. Traditional models based on Probabilistic Graphical Models (PGM) interpret interactions (e.g., clicks) between users and search engines as implicit relevance feedback, describing user behavior through a series of observable and latent events\cite{DBLP:conf/wsdm/CraswellZTR08, DBLP:conf/wsdm/GuoLW09, DBLP:conf/www/ChapelleZ09, DBLP:conf/sigir/DupretP08}. Borisov et al. proposed the Neural Click Model (NCM) \cite{DBLP:conf/www/BorisovMRS16}, in which queries, user interactions, and results are vectorized, thereby enhancing the predictive capability beyond traditional click models. Regarding user stopping behavior, programmable simulation stopping strategies primarily utilize heuristic methods based on judgment rules\cite{DBLP:journals/jasis/Cooper73a, DBLP:journals/ipm/KraftL79, 1995Judgment, DBLP:conf/cikm/MaxwellAJK15, DBLP:journals/sigir/Maxwell19}.

\noindent \textbf{LLM-powered agents.} Numerous studies aim to enhance LLMs' capabilities in executing complex tasks\cite{DBLP:conf/nips/Wei0SBIXCLZ22, DBLP:conf/iclr/YaoZYDSN023, DBLP:journals/corr/abs-2303-17651, shinn2023reflexion, gong2024cosearchagent}. Furthermore, the reasoning abilities of LLMs have been expanded through practical integration with external tools, such as APIs, search engines, calculators, or other models\cite{DBLP:journals/corr/abs-2302-04761, DBLP:journals/corr/abs-2303-17580, DBLP:journals/corr/abs-2303-08128}. Recently, some research has attempted to utilize LLMs to generate simulated queries\cite{DBLP:journals/corr/abs-2303-07678, DBLP:conf/sigir/MackieC023, DBLP:journals/corr/abs-2308-00415}. Engelmann et al. highlighted that current user models overlook the context and introduced the simulation of context-driven query reformulations\cite{DBLP:journals/corr/abs-2312-09631}. Although this approach enables models to acquire more relevant information by considering contextual data, existing efforts are still primarily limited to the level of generating single queries. These studies have not fully leveraged the multitasking capabilities of LLMs in dealing with broader, session-level tasks, nor have they effectively utilized the reasoning abilities of LLMs to simulate the coherent behavior of users throughout the entire search session.

\vspace{-8pt}
\section{METHODOLOGY}

\begin{figure*}[t]

  \centering
  \setlength{\abovecaptionskip}{0.2cm}
  \includegraphics[width=0.85\textwidth]{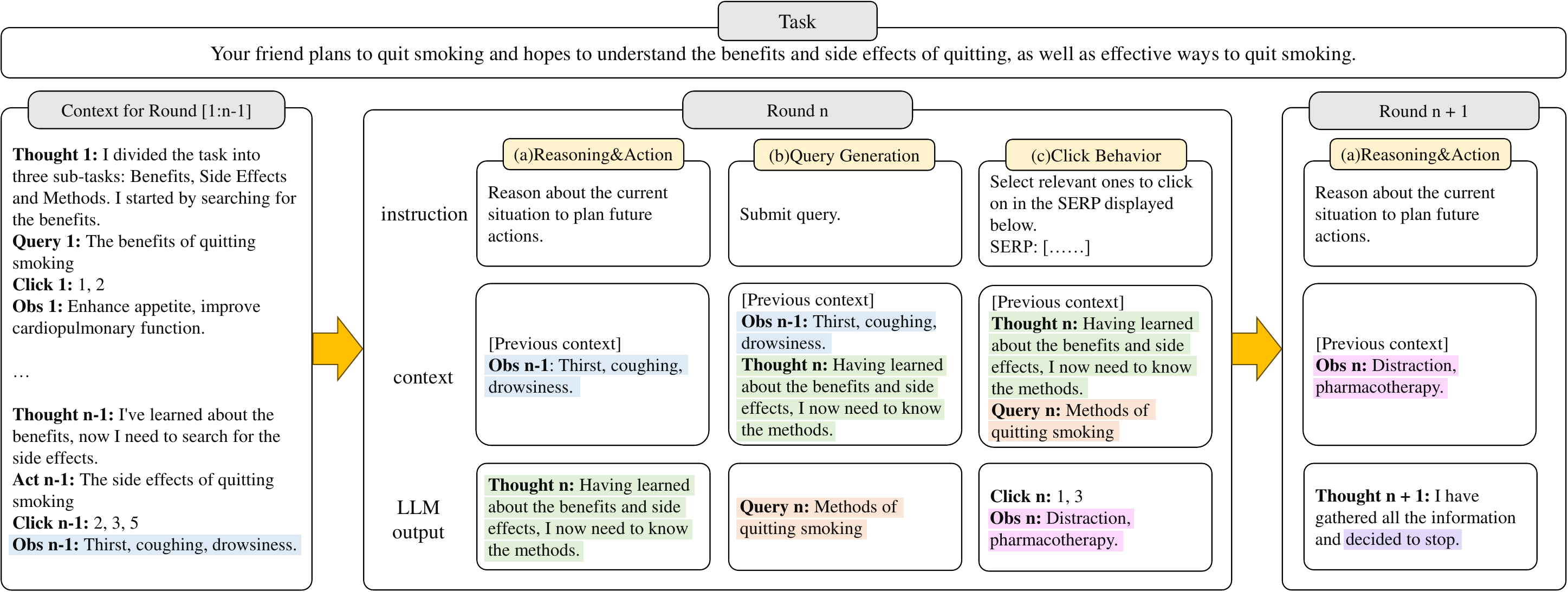}
  \caption{The overall framework of USimAgent. Given task T, the USimAgent will generate a search interaction sequence. USimAgent operate in multiple rounds, each consisting of three different steps, all executed by an LLM: (a) Reasoning\&Action; (b) Query Generation; (c) Click Behavior. In each step, the USimAgent takes into account the context generated from previous search interactions.}
  \label{fig:Method}
  \vspace{-12pt}
\end{figure*}

\noindent \textbf{Problem formulation.} Given a task $T$, our goal is to generate a search interaction sequence $I=(q_1, c_1, \ldots, q_n, c_n, s)$, which consists of alternating queries $q$ and clicks $c$, and concludes with a session stop action $s$. We divide the complete interaction sequence into multiple rounds, each round comprising the decision to stop, along with the actions of querying and clicking. This section first discusses how to generate actions containing reasoning under the guidance of context, and then explores the specific implementation processes of query construction and click behavior.

\noindent \textbf{Reasoning before acting.} To enhance the accuracy of actions generated by the USimAgent, we refer to the ReAct method\cite{DBLP:conf/iclr/YaoZYDSN023}, expanding the action space to a combination space of possible reasoning and action steps. Specifically, the action space $\hat{A}=A\cup L$, where $L$ is the language space. At round $t$, we have the context of the previous $t-1$ rounds, denoted as $C_t=(r_1, q_1, o_1 ,... ,r_{t-1}, q_{t-1}, o_{t-1})$, where $o$ denotes the observed information and $r$ represents reasoning. Based on this, USimAgent generates reasoning $r_t=LLM(\text{Prompt}_r(T,C_t))$ tailored to the task description and current context, and updates the context $C_t\leftarrow(C_t, r_t)$ accordingly. Here, $\text{Prompt}_r$ is a reasoning prompt, guiding the LLM to perform specific reasoning (the same applies for $\text{Prompt}_r$  with subscripts s, q, c, o, which respectively guide the LLM to generate stops, queries, clicks and observations). Then the agent must make a decision $s_t=LLM(\text{Prompt}_s(T,C_t))$ on whether to terminate the search session based on the reasoning. Depending on $s_t$, $I$ will be updated differently.

\vspace{-10pt}
\begin{equation}
\
I\leftarrow
\begin{cases} 
I & \text{if } s_t = \text{continue} \\
(I, s) & \text{if } s_t = \text{stop}
\end{cases}
\vspace{-5pt}
\end{equation}

When $s_t =  \text{continue}$, it moves to the query reformulation step; otherwise, the session ends, and the action sequence $I$ is output.

\noindent \textbf{Query reformulation.} In the query reformulation step, it generates a query $q_t=LLM(\text{Prompt}_q(T,C_t))$ and updates the context $C_t\leftarrow(C_t, q_t)$ and the interaction sequence $I \leftarrow (I, q_t)$.

\noindent \textbf{Click Prediction.} The agent receives the SERP for the query $q_t$ , denoted as $\text{SERP}_q$, which is a list that includes the title and abstract for each result. Based on task relevance, the agent selects results $c_t=LLM(\text{Prompt}_c(T,C_t))$ to click and updates the interaction sequence $I\leftarrow(I, c_t)$. Following this, by reading the detailed content of the clicked results, the agent acquires observations $o_t=LLM(\text{Prompt}_o(T,C_t))$ and updates the context $C_{t+1}=(C_t, o_t)$, proceeding to a new round of reasoning.

\vspace{-5pt}
\section{Experiments}

\subsection{Datasets and Experimental Settings}

\subsubsection{Datasets}
We evaluated our USimAgent on public user behavior dataset\footnote{http://www.thuir.cn/KDD19-UserStudyDataset/}, which was collated by Liu et al. through a laboratory user study\cite{DBLP:conf/kdd/LiuMLZM19}. Participants completed nine intricate search tasks, with the dataset incorporating their search interactions, which include query reformulations, clicks, and stopping behaviors.

We adhere to several rules for data filtration:

\begin{itemize}
\item Due to various reasons, search URLs for four tasks were predominantly invalid, unable to replicate the content originally viewed by the users, hence, those tasks were excluded.
\item The original dataset featured SERPs of varying lengths; for lists exceeding ten results, only the top ten were retained.
\item Incomplete sessions, characterized by missing click information or having partial SERPs, were removed.
\end{itemize}

Finally, 164 sessions from 40 users met these constrictions, and we retained these search sessions to form the evaluation dataset.

\subsubsection{Baselines and Evaluation Metrics}

We delineate the session generation process into three distinct phases: query reformulation, clicks and stopping behaviors, and choose baseline models for each phases for a comparative analysis with the proposed USimAgent.

\noindent \textbf{Query reformulation.} Following Azzopardi et al.\cite{DBLP:conf/sigir/AzzopardiRB07, DBLP:conf/sigir/Azzopardi09}, we utilize a generative probabilistic model for term sampling within a language model, which is composed of documents and task descriptions. This language model is constructed by linearly weighting four components: task descriptions, document titles, abstracts, and body texts, with the adjustment of weights aimed at optimizing model performance. The length of the queries follows a Poisson distribution, with its mean set to the average length of actual queries. We implemented two user querying strategies: {\itshape Random Selection} and {\itshape Popular Selection}, with specific details of these strategies following the methodology described in \cite{DBLP:conf/sigir/AzzopardiRB07}.

\noindent \textbf{Clicks.} We compare our USimAgent  against two categories of click models: (1) traditional probabilistic graphical models, including Position-Based Model (PBM)~\cite{DBLP:conf/wsdm/CraswellZTR08}, User Browsing Model (UBM)~\cite{DBLP:conf/sigir/DupretP08}, Dependent Click Model (DCM)~\cite{DBLP:conf/wsdm/GuoLW09}, and Dynamic Bayesian Network Model (DBN)~\cite{DBLP:conf/www/ChapelleZ09}, implemented using the open-source Py-Click\footnote{https://github.com/markovi/PyClick}, and (2) the neural click model NCM\footnote{https://github.com/CHIANGEL/Neural-Click-Model}\cite{DBLP:conf/www/BorisovMRS16}. Given that the training of click models relies on extensive click data, and our session data is relatively limited, to ensure fairness in comparison, we attempt to address the issue of data sparsity in the training process of the PBM by employing a regression-based EM algorithm\cite{10.1145/3159652.3159732}.

\noindent \textbf{Stopping behavior.} We replicated several existing stopping strategies, including Fixed Depth\cite{balog2023user}, Frustration Point\cite{DBLP:journals/jasis/Cooper73a, DBLP:journals/ipm/KraftL79}, Satisfaction Point\cite{DBLP:journals/jasis/Cooper73a, DBLP:journals/ipm/KraftL79}, and a combination of Frustration and Satisfaction Points\cite{DBLP:journals/ipm/KraftL79}, using these as baselines for comparison.

For query reformulation, we utilized BLEU\cite{DBLP:conf/acl/PapineniRWZ02}, which employs heuristic rules such as n-gram matching to measure the similarity between simulated and real queries. For clicks and stopping behaviors, we assessed performance based on accuracy, precision, recall, and the F1 score.

\subsubsection{Implementation Details}

Our system employs GPT-4\footnote{https://chat.openai.com/}, a widely recognized and extensively utilized LLM, with the configuration settings of “temperature = 0, n = 1”.

\vspace{-5pt}
\subsection{Results and Analysis}

\begin{table}
\setlength{\abovecaptionskip}{0.2cm}
  \caption{Similarity between query reformulation strategies and real queries. We highlight the highest scores in bold.}
  \label{tab:query}
  \begin{tabular}{lc}
    \toprule
    Metric&BLEU\\
    \midrule
    Random Selection&0.1417\\
    Popular Selection&0.2765\\
    USimAgent&\textbf{0.4630}\\
  \bottomrule
\end{tabular}
\vspace{-15pt}
\end{table}

\subsubsection{Main Results}

Table \ref{tab:query} presents the similarity between the generated queries and the real user queries on the user behavior dataset for both baselines and USimAgent. The Random Selection method assigns the same sampling probability to all words in the corpus, neglecting the differences in importance among words, while the Popular Selection method considers the weight of words, resulting in a higher BLEU score. USimAgent achieved the best results in the experiments, significantly outperforming the baseline methods, which validates the effectiveness of our approach.

\begin{table}
\setlength{\abovecaptionskip}{0.2cm}
  \caption{Performance comparison of click models and USimAgent in user click behavior prediction("PBM-R" denotes the PBM trained using a regression-based EM algorithm). During the training of our click models, we employed a 10-fold cross-validation technique to partition the dataset into training and testing sets. We highlight the highest scores in bold.}
  \label{tab:click}
  \begin{tabular}{lcccc}
    \toprule
    Metric&Accuracy&Precision&Recall&F1 Score\\
    \midrule
    PBM&0.8082&0.7605&0.5131&0.6040\\
    UBM&\textbf{0.8174}&0.7735&0.5424&\textbf{0.6321}\\
    DBN&0.8053&\textbf{0.8053}&0.4737&0.5893\\
    DCM&0.7823&0.7640&0.4152&0.5292\\
    PBM-R&0.7583&0.6044&0.5302&0.5575\\
    NCM&0.7348&0.5061&0.5433&0.5175\\
    USimAgent&0.7458&0.4912&\textbf{0.6022}&0.5411\\
  \bottomrule
\end{tabular}

\vspace{-5pt}
\end{table}

Table \ref{tab:click} shows the performance of various click models compared to our USimAgent in predicting user clicks. UBM demonstrating better predictive performance. Applying the regression EM algorithm to PBM does not improve the predictive performance of PBM, and the poor performance of NCM may be due to underfitting caused by a sparse dataset.  After integrating all metrics, the performance of our USimAgent is comparable to that of traditional models. Potential reasons for this compromised include: (1) Accounting for position bias is crucial in click prediction models. USimAgent's failure to effectively capture position bias impacts its performance, whereas classic click models such as UBM and PBM possess more sophisticated mechanisms for addressing position bias. (2) USimAgent employs a zero-shot learning approach, which, despite its utility in scenarios of data scarcity, may fall short in predictive performance compared to models trained on large-scale datasets.

\begin{table}
\setlength{\abovecaptionskip}{0.2cm}
  \caption{Performance comparison of various stopping strategies and USimAgent in user stop behavior prediction. We highlight the highest scores in bold.}
  \label{tab:stop}
  \begin{tabular}{lcccc}
    \toprule
    Metric&Accuracy&Precision&Recall&F1 Score\\
    \midrule
    Fixed depth&0.6679&0.4541&\textbf{0.6341}&\textbf{0.5293}\\
    Satisfaction&0.7361&0.5780&0.3841&0.4615\\
    Frustration&0.7038&0.4959&0.3720&0.4251\\
    F\&S&0.7145&0.5150&0.5244&0.5196\\
    USimAgent&\textbf{0.8043}&\textbf{0.9365}&0.3598&0.5198\\
  \bottomrule
\end{tabular}
\vspace{-10pt}
\end{table}

In the task of predicting stopping behavior, as shown in Table \ref{tab:stop}, USimAgent demonstrates superior performance in Accuracy and Precision but does not surpass the baseline methods in Recall and F1 Score.

\subsubsection{Ablation Study}

\begin{figure}[t]
\setlength{\abovecaptionskip}{0.2cm}
  \centering
  \vspace{-5pt}
  \includegraphics[width=0.7\linewidth]{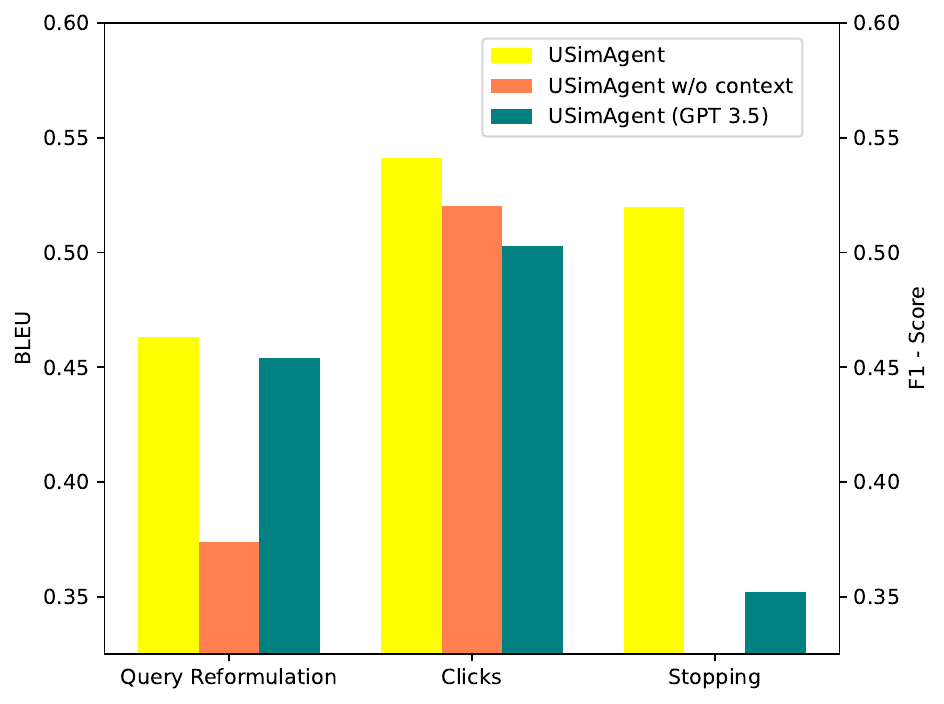}
  \caption{Ablation results. For the prediction of clicks and stopping behavior, we employ the F1 Score as the evaluation metric.}
  \label{fig:Ablation}
  \vspace{-15pt}
\end{figure}

To validate the significance of context information for generating complete sessions, we present the results of an ablation study in Figure \ref{fig:Ablation}. Additionally, we constructed USimAgent using the most advanced LLM available and explored whether this framework could be effectively transferred to other, less powerful LLMs. Specifically, we employed GPT-3.5 to assess the performance of USimAgent with less powerful LLMs. In the task of query generation, the removal of context information resulted in USimAgent's inability to effectively decompose complex tasks into multiple subtasks, leading to a significant decline in query generation capabilities. Furthermore, USimAgent was unable to accurately determine whether the information collected was sufficient to complete the task, thereby impeding its ability to make reasoned decisions about when to cease querying. Comparatively, the impact of context information on click prediction tasks was minor, likely because click tasks are more independent and rely less on historical search information.

Although applying USimAgent to GPT-3.5 resulted in a decrease in performance, it still demonstrated superior capabilities compared to removing context information. This indicates that our proposed framework is capable of effectively learning and utilizing context. With the development of more powerful models in the future, USimAgent's performance in such tasks is expected to see further enhancement.

\vspace{-5pt}
\section{Conclusion and future work}

In this paper, we introduced USimAgent, a new framework leveraging LLMs for search user simulation. Empirical investigation on a real user behavior dataset shows that the proposed simulator outperforms existing methods in query generation and is comparable to traditional methods in predicting user clicks and stopping behaviors. Although the LLM-based USimAgent demonstrates promising capabilities in zero-shot scenarios, its predictive accuracy may still fall short compared to models trained on more extensive datasets. Therefore, combining LLMs with a broader dataset could represent a future research direction for simulating user searches.


\vspace{-5pt}
\begin{acks}
This research was supported by the Natural Science Foundation of China (61902209, 62377044, U2001212), and Beijing Outstanding Young Scientist Program (NO. BJJWZYJH012019100020098), Intelligent Social Governance Platform, Major Innovation \& Planning Interdisciplinary Platform for the "Double-First Class" Initiative, Renmin University of China.
\end{acks}

\bibliographystyle{ACM-Reference-Format}
\bibliography{sample-base}

\end{document}